\documentstyle[psfig,prb,aps]{revtex}
\begin{document}

\headheight-1.0cm
\headsep0.0cm
\topmargin0.6cm
\textheight24cm

\draft

\twocolumn[\hsize\textwidth\columnwidth\hsize\csname
@twocolumnfalse\endcsname

\title{Theoretical analysis of the electronic structure 
of the stable and metastable $c(2 \times 2)$ phases of Na on Al\,(001):\\
Comparison with angle-resolved ultra-violet 
photoemission spectra}
\author{C. Stampfl$^{a}$, K. Kambe$^{b}$, R. Fasel$^{c}$, 
 P. Aebi$^{c}$, and M. Scheffler$^{d}$}
\address{$^{a}$ Xerox Palo Alto Research Center, 3333 Coyote Hill Road, 
Palo Alto, California  94304\\ 
$^{b}$ Prinzregentenstrasse 57/59, D-10715 Berlin, Germany\\
$^{c}$ Institut de Physique, Universit\'{e} de Fribourg, Peroll\'{e}s, 
1700 Fribourg, Switzerland\\
$^{d}$ Fritz-Haber-Institut der Max-Planck-Gesellschaft, Faradayweg 4-6, D-14195
Berlin-Dahlem, Germany
}
\maketitle

\begin{abstract}
Using Kohn-Sham wave functions and their energy levels obtained by 
density-functional-theory total-energy calculations,
the electronic structure of the 
two $c(2 \times 2)$ phases of Na on Al\,(001) are analysed;
namely, the metastable hollow-site structure formed when adsorption takes place 
at low temperature, and the
stable substitutional structure appearing when the substrate is heated
thereafter above ca. 180K or when adsorption takes place at room temperature 
from the beginning.  The experimentally obtained two-dimensional 
band structures of the surface states or resonances are well reproduced by 
the calculations.
With the help of charge density maps it is found that in both 
phases, two pronounced bands appear as the  result of 
a characteristic coupling between the 
valence-state band of a free $c(2 \times 2)$-Na monolayer and the 
surface-state/resonance band of the Al surfaces; that is, the clean (001) 
surface for the metastable phase  and the unstable, reconstructed 
``vacancy'' structure for the stable phase. 
The higher-lying band, being Na-derived, remains metallic for the unstable phase,
whereas it lies completely above the Fermi level for the stable phase, leading to 
the formation of a surface-state/resonance band-structure resembling the bulk 
band-structure of an ionic crystal.
\end{abstract}

\vskip2pc]

\section{Introduction}

The adsorption of alkali metal atoms on metal surfaces has attracted
much attention in recent years partly due to the
discovery of a variety of new
adsorbate phases, in particular, structures that involve
a reconstruction of the metal
surface induced by the alkali metal atoms (see, for example, 
Refs.~\cite{reviews,renee}
and references therein). A common feature of these
systems is that often there is a metastable phase at low 
temperature involving no reconstruction of the metal surface
while at higher temperatures,
the stable reconstructed phase occurs.

In the present paper we present a combined theoretical and experimental
investigation of one such system, namely
that of $c(2 \times 2)$ phases of Na on Al\,(001).  
Early dynamical-theory analyses of
low energy electron diffraction (LEED) intensities
\cite{hutchins,vanhove} concluded that Na occupied the four-fold hollow site.
It was first demonstrated by high resolution core-level spectroscopy
(HRCLS) \cite{andersen1} and surface extended x-ray absorption fine structure
(SEXAFS) \cite{aminpirooz} studies that the hollow site is taken only for 
preparations at low temperature (LT), and a different, stable structure is formed
by adsorption at room temperature (RT) or by heating of the LT phase above ca.
 160K.
It was shown later by a density functional theory (DFT) study \cite{stampfl}
and by a new LEED analysis \cite{berndt} that in the RT phase the Na 
atoms occupy substitutional sites, where every second Al atom
in the top layer is displaced and a Na atom adsorbed in its place.
The result of an x-ray photoelectron diffraction (XPD)
study \cite{roman} of the RT phase concluding that the surface
contains two domains, could not be confirmed \cite{berndt}.

In what follows, we perform a theoretical analysis of the {\em electronic}
structure of the  two  $c(2 \times 2)$   structures and compare the results with
 those of angle-resolved ultraviolet spectroscopy  
\cite{fasel}. The measurements have been performed in what we
call ``polar scan'' modes
 which deliver displays of photoemission intensities
as a function of energy and  wave vector component parallel to
the surface, lying in selected symmetry directions 
($\overline{\Gamma}$-$\overline{\rm M}$ and $\overline{\Gamma}$-$\overline{\rm X}$)
in the two-dimensional (2D) Brillouin zone (see Fig.~1). The displays yield 
directly 2D band structures of surface 
states/resonances (i.e. surface states {\em or} resonances,
depending on the position of the states in or out of the gap of the 2D projected
bulk bands), which may be compared with 
calculated results.

The basis of our theoretical
analysis are Kohn-Sham wave functions and their energy levels obtained by 
DFT total energy
calculations \cite{stampfl}. 2D band structures are derived and 
compared with the experimental results.
The obtained, satisfactory agreement between theory and experiment
may be regarded as a confirmation of the proposed atomic structure models 
mentioned above, and a useful basis for further studies of the properties of
these surfaces.

Also single-state charge density distributions of occupied and unoccupied
states are derived from the DFT calculation, and used for analysing the 
character of the bands.
We find that adsorption leads, in both LT and RT cases, to {\em two} main bands
of surface state/resonances as the  result of a coupling between the
valence-state band of a free $c(2 \times 2)$-Na monolayer and the 
surface-state/resonance band of the Al surface. Across the Fermi level the 
Al-derived band is shifted 
down, and the Na-derived band is shifted up. As a consequence, a
charge transfer is taking place from the adsorbate directly into the 
{\em pre-existent surface state/resonance} of the substrate. 
%
%
\begin{figure}
\vspace{-10mm}
\psfig{figure=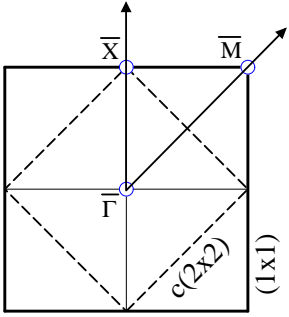,rheight=45mm}
\caption{Surface Brillouin zone of $(1 \times 1)$ (full lines)
and $c(2 \times 2)$ (broken lines) structures. The symmetry points
$\overline{\Gamma}$, $\overline{\rm M}$, and $\overline{\rm X}$
are referred to the $(1 \times 1)$ structure.}
\end{figure}
For the LT (hollow) structure, it is found that the coupling is relatively weak 
and the Al-derived state almost retains the perfect $(1 \times 1)$
periodicity of the clean surface. For the RT (substitutional)
structure, the Al-derived state exhibits a clear 
$c(2 \times 2)$ periodicity
of the reconstructed Al surface. The Na-derived band of the RT structure 
is completely empty and leads, except for the still 
existent background bulk-continuum, to a 
surface electronic structure having a character of an {\em ionic}
monolayer lying on the surface. We discuss possible 
consequences of this finding.

The paper is organised as follows: In Section~II the experimental method
is outlined and is followed in Section~III by a description
of the calculation methods.  Section~IV contains
the results for the surface state/resonance band structures
and the analysis of the character
of the bands in terms of single-state charge density
distributions. Also the charge transfer processes are analysed by using 
density differences.
Section~V contains the discussion and Section~VI, the
conclusion.

\section{Experiment}

The photoemission experiments \cite{fasel} were performed in a VG ESCALAB Mark II
spectrometer modified for motorized sequential angle-scanning data acquisition
\cite{osterwalder},
 and with a working pressure in the lower 10$^{-11}$ mbar region. Ultraviolet
photoelectron spectroscopy (UPS) measurements were done using unmonochromatized
He I (21.2~eV) radiation from a discharge lamp. The 150-mm radius hemispherical
analyzer was run with an energy resolution of 50 meV. Contamination free
surfaces were prepared by a combination of Ar$^{+}$ sputtering and annealing at 500
C. Na was evaporated from a carefully outgassed SAES getter source. Particular
care was taken to ensure ultraclean Na deposits: All parts of the evaporation
source, except the tiny exit slit, were surrounded with liquid-nitrogen cooled
walls. In this way, the pressure during evaporation could be kept as low as
2.5$\times$10$^{-11}$ mbar. Na coverages were determined accurately 
($\pm 0.03$ ML) from
core-level photoelectron intensities \cite{osterwalder2}.
 The sample temperature was measured
with a thermocouple in mechanical contact with the sample holder. The
temperature gradient between the sample and the sample holder was determined in
a separate calibration experiment with an additional thermocouple spot-welded
onto a dummy sample. Sample temperatures given here are corrected for this
temperature difference and are estimated to be correct within $\pm$10~ K.

In order to obtain two-dimensional band structures of surface states/resonances,
polar scans along the $\overline{\Gamma}$-$\overline{\rm M}$ and 
$\overline{\Gamma}$-$\overline{\rm X}$ directions of the
 Al\,(001)
surface Brillouin zone were performed, recording at each angular setting
($\Delta \Theta = 2^{\circ}$) the entire photoelectron spectrum 
between $-$0.4 and 4.3~eV
binding energy. The experimental data sets I($E_{i},\Theta$)
 acquired in this way
were mapped onto a regular ($E_{i},{\bf k}_{\parallel}$)
 grid and visualized as gray-scale plots
with low intensities in white and high intensities in black.

\section{Calculation Method}

The {\em ab initio} DFT total energy calculations, and comparison with LEED
results are described in Refs.~\cite{stampfl,berndt}. 
The calculations were performed using the local-density approximation (LDA) for
the exchange-correlation functional.
Further details about the method can be found in 
Refs.~\cite{stampfl,neugebauer,stumpf}.

The following two kinds of diagrams have been used in the present analysis:

\begin{enumerate}
\item{\em 2D band structures}

For deriving 2D band structures we use the following procedure:\\
The projected density of states (DOS) onto a chosen 
atomic orbital, $\psi_{{\rm AO}}({\bf r})$, is defined as
\begin{equation}
N_{{\rm AO}}(\varepsilon) = \sum_{{\bf k}_{\parallel}} N_{{\rm AO}}
({\bf k}_{\parallel},
\varepsilon)  ,
\label{eq1}
\end{equation}
where the ``${\bf k}_{\parallel}$-resolved'' projected DOS is given
by 
\begin{displaymath}
N_{{\rm AO}}({\bf k}_{\parallel},\varepsilon)=\sum_{\varepsilon'}
\vert \int_{r<r_c} d^{3}{\bf r} \psi_{{\rm AO}}^{*}({\bf r})
 \psi_{{\bf k}_{\parallel}, \, \varepsilon'}
({\bf r})\vert^{2}
\end{displaymath}
\begin{equation}
\times \frac{\gamma}{\pi}\frac{1}{(\varepsilon' - \varepsilon)^{2} + \gamma^{2}}
\quad  .
\label{eq3}
\end{equation}
We have used for $\psi_{\rm AO}(\bf r)$ the eigenfunctions of the isolated 
pseudo-atoms from which the pseudopotentials were derived for the total energy 
calculation~\cite{stampfl}. 
The integral was truncated at a cutoff radius $r_{c}$. (We used here
$r_{c}=3.7$\AA\,.)
In the standard supercell method using a slab geometry, 
a Kohn-Sham state,  $\psi_{{\bf k}_{\parallel}, \, \varepsilon}
({\bf r})$, can be 
specified, except for the presence of degeneracy,
by indices ${\bf k}_{\parallel}$
and $\varepsilon$ (both discrete), the parallel wave vector and the energy,
respectively.
The lifetime broadening parameter, $\gamma$, is introduced for
convenience in numerical work and for improving visibility of peaks
in the resulting plots. (We take here $\gamma$=0.5~eV.)

A simple sum of $N_{{\rm AO}}({\bf k}_{\parallel},\varepsilon)$'s
are formed over atomic orbitals of a specified atom. For example, for Na we have,
\begin{displaymath}
N_{\rm Na}({\bf k}_{\parallel}, \varepsilon) = N_{{\rm Na}\, 3s}
({\bf k}_{\parallel},
\varepsilon) + N_{{\rm Na}\, 3p_{x}}({\bf k}_{\parallel}, \varepsilon) 
\end{displaymath}
\begin{equation}
+  N_{{\rm Na}\, 3p_{y}}({\bf k}_{\parallel}, \varepsilon) +
N_{{\rm Na}\, 3p_{z}}({\bf k}_{\parallel}, \varepsilon)
\quad  .
\label{eq5}
\end{equation}
The corresponding quantities, 
$N_{{\rm Al}}({\bf k}_{\parallel},\varepsilon)$, are 
evaluated for the Al atoms in the uppermost layer.
The maximum peak of these quantities was always found at the bottom
of a surface state/resonance band lying at $\overline{\Gamma}$. Those peaks
having a fractional ratio to the maximum higher than a properly 
chosen fixed value (``Min. Fraction'' in Tab.~I),
 are selected and their positions in 
$({\bf k}_{\parallel}, \varepsilon)$-
space displayed, using squares for Na and 
circles for Al (see Tab.~I). 
The plots along $\overline{\rm X}$-$\overline{\Gamma}$ 
and $\overline{\Gamma}$-$\overline{\rm M}$ (see Fig.~1) are combined, and   
are presented in Figs.~2 and 3. We see that they produce 
satisfactorily 2D band structures to be compared to the
experimentally obtained ones.   

The bulk-band continuum is separated into discrete bands due to the use of finite
(nine layer) Al slabs. They appear in Figs.~2 and 3 as weak features.
The discreteness of bulk bands may have caused a small energy shift
of the surface {\em resonance} bands lying inside the bulk continuum, because 
these are replaced by one of the discrete bulk bands lying nearest to them.

We note that in Figs.~2 and 3 the theoretical bands for the LT and RT structures
repeat themselves inside the $(1 \times 1)$ Brillouin zone, exhibiting
the $c(2 \times 2)$ periodicity.
That is, the range $\overline{\Gamma}$-$\overline{\rm M}$
($\overline{\rm M}$ being referred to $(1 \times 1)$) is halved, and the bands 
in the second half become a mirror image of the first half.  We call it here 
"back-folding" (cf. Ref.\cite{hora}). It is a 
consequence of our use of density of states projected to one specified atom
in each $c(2 \times 2)$ surface unit cell.
We note that this is obviously a theoretical construct and not quite 
adequate for representing fully the character of the wave functions of the 
bands, particularly in the LT structure.  In fact, there is no information 
included about the relation between the values of wave functions around the two 
Al atoms in a unit cell (see Fig.~4).
Thus, in reality the wave functions may happen to have approximately the 
$(1 \times 1)$ Bloch-type periodicity -- the back-folding still occurs.
The same would also result even for the clean-surface band if 
the artificial $c(2 \times 2)$ unit cell would be imposed.
In our back-folded band structure, a vanishing
deviation from the $(1 \times 1)$ Bloch-type periodicity 
would become visible only in a vanishing band gap at a new Brillouin zone boundary, 
that is,
in our case on the line halving the range $\overline{\Gamma}$-$\overline{\rm M}$. 

On the other hand, the photoemission intensity would be determined by a matrix 
element
\begin{equation}
M({\bf k}_{\parallel},\varepsilon)=
 \int d^{3}{\bf r} \psi_{{\rm f}}^{*}({\bf r})\nabla
 \psi_{{\bf k}_{\parallel}, \, \varepsilon'}
({\bf r})  .
\label{eq6}
\end{equation}
The symmetry of the structure leads to
the "selection rule", which selects out initial-state bands according to the 
symmetry relation between the initial- and final-state wave functions.
In the case of the LT structure, the final state $\psi_{{\rm f}}({\bf r})$
samples the initial state $\psi_{{\bf k}_{\parallel}, \, \varepsilon'}({\bf r})$
at both of the two Al atoms in a unit cell.
The selection rule can take place differently for the first 
and second halves of the range
$\overline{\Gamma}$-$\overline{\rm M}$, destroying the mirror symmetry 
between the two halves. In particular, if the wave functions have 
nearly the $(1 \times 1)$ Bloch-type periodicity, the selection rule leads to 
"unfolding" of the back-folded band structure.

\item{\em Charge density distributions}

The charge density is derived from Kohn-Sham wave functions as
\begin{equation}
\rho({\bf r}) = \sum_{{\bf k}_{\parallel}}
 \sum_{\varepsilon \, {\rm occ}}
\vert \psi_{{\bf k}_{\parallel}, \varepsilon}  ({\bf r})\vert^{2}  ,
\label{eq7}
\end{equation}
where $\varepsilon$ occ labels the 
occupied levels. We use $\rho({\bf r})$ for the study of charge transfer ocurring 
at adsorption.

 The single-state charge density is defined here as
\begin{equation}
\rho_{{\bf k}_{\parallel}, \varepsilon}({\bf r}) =
 \vert \psi_{{\bf k}_{\parallel}, \varepsilon}  ({\bf r})\vert^{2}
\quad  .
\label{eq9}
\end{equation}
and used for analysing the character of surface states/resonances.

We should keep in mind that the charge density distributions
are derived from {\em pseudo}-wave-functions resulting from the use of
pseudopotentials, so that they give a correct distribution  
only outside the critical radius of the pseudopotentials.

We use cross-sections of the charge densities on planes
perpendicular and parallel to the surface.
The perpendicular cross-sections taken are specified in Fig.~4.

\end{enumerate}

\section{Results}

In the next three Sections (A, B and C) we 
discuss successively the results for the clean Al\,(001) and the two
$c(2 \times 2)$-Na/Al\,(001) surfaces. It is to be noted that the Figures are
arranged, independently of the text, in the way to facilitate a visual 
comparison between the different structures.
\twocolumn[\hsize\textwidth\columnwidth\hsize\csname
@twocolumnfalse\endcsname
%
%
\begin{figure}
\vspace{-10mm}
\psfig{figure=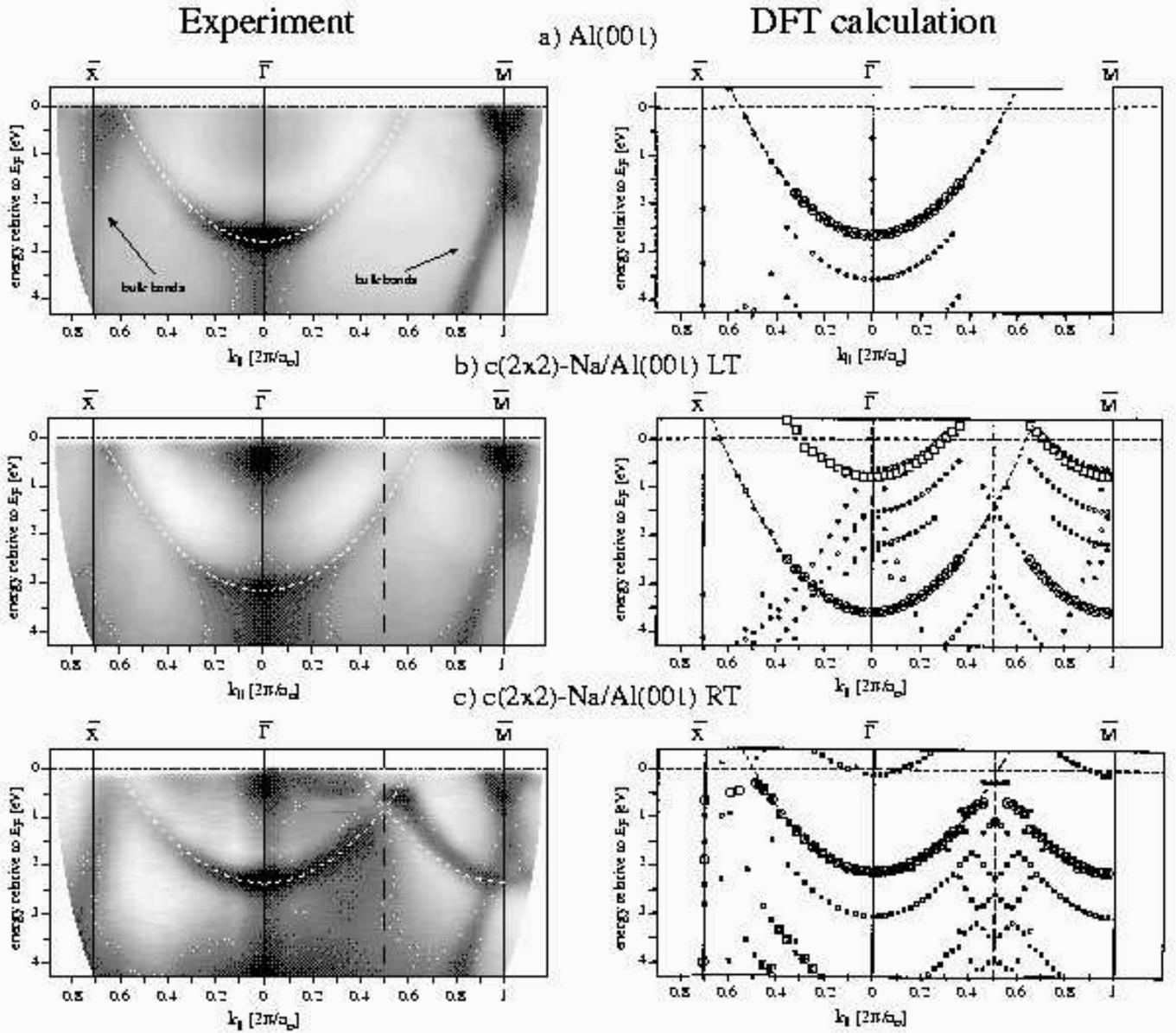,width=180mm,rheight=165mm}
\caption{Comparison of experimental (left panel) and calculated (right panel)
2D band structures.
The symbols used for the calculated bands are specified in Tab.~I.
(a) Clean surface, (b) LT phase (hollow site), and (c) RT phase (substitutional
site).}
\end{figure}
\vskip2pc]

\subsection{Clean Al\,(001)}
Figure~2a shows the experimental (left panel)  and theoretical
(right panel) band structures for the clean Al\,(001) surface.
It can be seen that both exhibit a surface-state band
which has a free-electron-like form. The paramaters obtained by fitting by 
parabolas, as indicated by broken lines in Figure~2a, are shown in Table II.
The energy position of the band at $\overline{\Gamma}$, $E_{0}$,
agrees well between theory and experiment, with values
of 2.68 and 2.76~eV below the Fermi level, respectively.
The theoretical value lies in the range of the results of other calculations
\cite{heinrichsmeyer}-\cite{caruthers}: 2.6 - 2.9 eV.
The experimental value is in very good agreement with earlier studies
 \cite{gartland}, \cite{hansson} and \cite{kevan}.

In Fig.~2 (left panel) it can be seen that the experimental results
display a number of additional features. Those which are common to all structures 
studied on this surface are assigned to bulk bands, 
as indicated by arrows.
These bulk features do not appear in the calculated band structures 
(right panel), reflecting the fact that the density of states projected 
onto the uppermost Al atoms are dominated by the surface states.  

In Fig.~6a we display the single-state charge density distributions
at $\overline{\Gamma}$ for the main band.  The two 
cross-sections (100) and (110) are defined in Fig.~4a.  An important
characteristic of the surface state of the
clean Al\,(001) surface is that the charge density shows a pronounced maximum
just on top of the surface Al atoms. This can 
also be seen in the cross-section parallel to  the surface through
the electron density maxima as shown in Fig.~7a. This
particular feature of the surface state is apparently crucial in the formation
of the electronic structure of the LT phase, as we see below.
%
%
\begin{figure}
\vspace{-10mm}
\psfig{figure=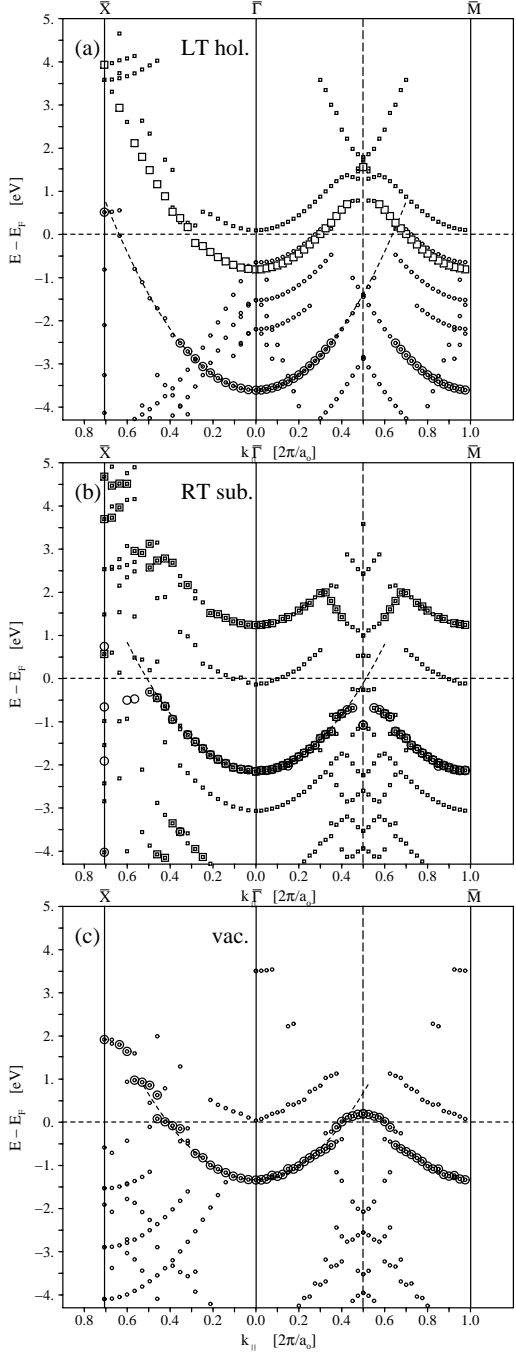,rheight=205mm}
\caption{2D band structures of (a) the LT phase (hollow site), (b) the
RT phase (substitutional site), and (c) the vacancy structure, including the
energy range above ${\rm E}_{\rm F}$. The symbols used are specified
in Tab.~I.}
\end{figure}

\subsection{LT $c(2 \times 2)$ hollow structure}

\subsubsection{Al-derived band}

In Fig.~2b we show the experimental (left panel)  and calculated
(right panel) surface state/resonances for the LT $c(2 \times 2)$ hollow 
structure. 
The lower lying main band is Al-derived
as indicated by the circles in the theoretical plot (cf. Table I). The position of
the calculated band at $\overline{\Gamma}$ lies somewhat lower in
energy than the experimental value; compare 3.61 to 3.12~eV (see Table II). 
(Notably the calculated results of the earlier theoretical work of 
Chulkov and Silkin \cite{chulkov} is also 
$\approx$3.6~eV, although their non-self-consistent value of vertical distance 
2.05 \AA, from Na to Al is different from ours 2.35 \AA, which is nearer to the
experimental result 2.57 A \cite{berndt}.)
Compared to the position of the surface-state band of the clean Al\,(001)
surface, the Al-derived band lies lower in energy, by 0.93~ and 0.36~eV,
as obtained by the calculations and as determined from experiment, respectively.
The nature of this downward shift will be discussed below.
The differences between theory and experiment may have been caused by the 
approximations used in the total energy calculation (use of LDA, Rydberg cut, 
k-point sampling etc. \cite{neugebauer,stumpf}).

The experimental results show clearly that the Al-derived state does {\em not} 
have the 
$c(2 \times 2)$ periodicity, but rather the $(1 \times 1)$ periodicity of the
clean surface. In fact, we find it a significant experimental
result that the main band in the LT phase has, throughout the whole
range of $\overline\Gamma$-$\overline{\rm M}$, almost the
same form as that of  the clean surface,  being only shifted down.
%
%
\begin{figure}
\vspace{-10mm}
\psfig{figure=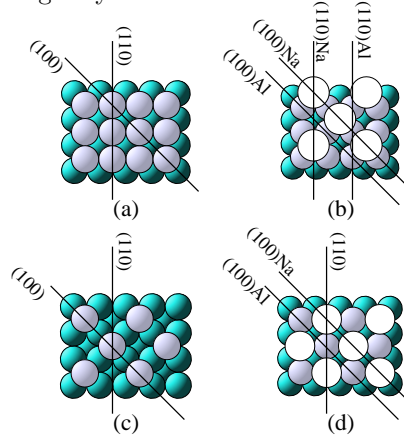,rheight=65mm}
\caption{Geometries of the $c(2 \times 2)$-Na/Al\,(001)
surface structures: (a) clean surface (b) LT phase (hollow site), (c)
vacancy structure and (d) RT phase (substitutional site).
The positions of perpendicular
cross-sections used for charge density distributions are indicated.
The white circles represent Na atoms and the grey circles Al atoms.}
\end{figure}

In Fig.~6b we display the single-state charge density distribution
at $\overline{\Gamma}$ for the Al-derived band of the LT phase.  The two sets of
cross-sections differently chosen for Al and Na (cf. Fig.~6c) are defined in 
Fig.~4b. It can
be noticed in Fig.~6b that the maxima of charge density remains on top of the 
surface 
Al atoms, almost unchanged from those of the clean surface shown in Fig.~6a.
(This has also been pointed out by Chulkov and Silkin~\cite{chulkov}.)
In the cross-section parallel to the surface shown in Fig.~7b only a small 
deviation from the $(1 \times 1)$ structure is seen, having a $c(2 \times 2)$ period,
with some indication of a character of a bonding state between
the Na and Al atoms.
The smallness of the deviation explains
the similarity of the experimental band structures between the
clean surface and the LT phase. It is to be noted that in the 
photoemission processes from this band, not only the initial
states, but also the final states have apparently maintained approximately
the same periodicity of the
clean surface. The theoretical curves in Fig.~2b appear different,
but this is only caused by the back-folding as already mentioned.

\subsubsection{Na-derived band}

In Fig.~2b dark spots can be seen at $\overline{\Gamma}$ and $\overline{\rm M}$,
near the Fermi energy, $E_{F}$, in the experimental results.  
Although from the figure presented here it is perhaps difficult
to distinguish, we find in the experimental data that the dark feature at  
$\overline{\rm M}$ for the
LT phase is markedly different in character from that of the clean surface,
in that the intensity for the LT phase is stronger and
some tailing with dispersion is exhibited, as is the case also for the feature at 
$\overline{\Gamma}$.
From the calculated bands of the LT phase (right panel of Fig.~2b),
we can see that there is a surface state/resonance band 
with an energy of about 0.7~eV
below the Fermi level at $\overline{\Gamma}$.
As indicated by the open squares, it is a Na-derived band. We 
assign this band as giving rise to the experimentally
measured intensities at $\overline{\Gamma}$
and $\overline{\rm M}$ at $\sim$0.4~eV below the Fermi level. 
The existence of this band was first theoretically predicted by Benesh {\em et al.}
\cite{benesh} and was also reproduced by Chulkov and Silkin \cite{chulkov}.
In Fig.~3a we show the same calculated band structure as in Fig.~2b, but where
the energy region extends higher into the positive range. 
Here we can see that the part of the Na-derived band above $E_{F}$
exhibits marked band-structure effects in the middle
of $\overline{\Gamma}$-$\overline{\rm M}$, due to the $c(2 \times 2)$ periodicity of 
the adsorbed Na layer.

In Fig.~5a we show the band structure of a free $c(2 \times 2)$-Na monolayer
for comparison. It can be observed that at $\overline{\Gamma}$ the lowest band 
lies $\approx$0.8~eV lower in energy as compared to the Na-derived band
of the LT phase. Correspondingly the occupied part of the band is notably
larger for the free monolayer. 

Figure~6c displays the distribution of the 
Na-derived surface state/resonance at 
$\overline{\Gamma}$ which clearly shows an anti-bonding character, the nodal
line (not shown) running between the Na- and Al-layers.
A comparison with Fig.~5b shows that the strongly smeared out character of the 
density between the Na atoms is maintained, appearing however only in the upper 
half of the Na layer. The lower half is apparently cancelled by the Al surface 
states due to the anti-bonding coupling.
A cross-section parallel to the surface passing through the electron
density maxima on top of the Na atoms is shown in Fig.~7c. We see also here 
a smeared out, free-electron-like distribution. 

\subsubsection{Formation mechanism of the two surface state/resonance bands}

Combining the results above, we can conclude that the two bands are resulting
from a coupling between the lowest lying (3$s$-derived)
band of a free $c(2 \times 2)$ monolayer of Na
and the surface state band of the clean Al\,(001) surface.
Applying a simple two-term perturbation theory,
as commonly done in molecular-orbital theory~\cite{coulson,atkins}, 
this may be understood as 
being due to the formation of bonding and
anti-bonding states, leading
to the downward shift of the Al-derived band by 0.9 eV (calc.) with an 
increase in population
and the upward shift of the Na-derived band by 0.8 eV (calc.) with a decrease
in population. Alternatively, it may be understood, as in the case of ionic 
crystals, that the shifts occur as the result of Coulomb fields between the 
two oppositely charged ionic layers.
The ionization is to be expected due to the cationic nature of Na, which
donates electronic charge to the aluminum.
%
%
\begin{figure}
\vspace{-10mm}
\psfig{figure=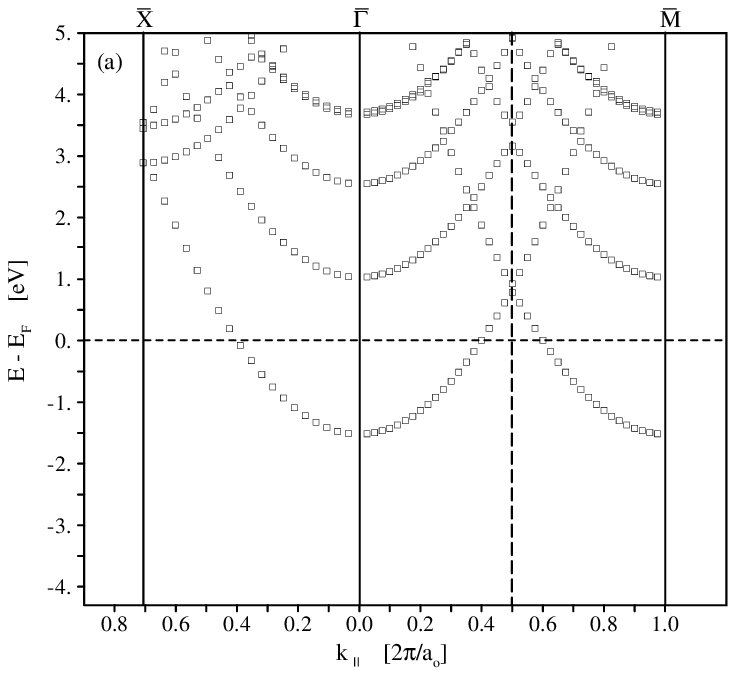,rheight=75mm}
\psfig{figure=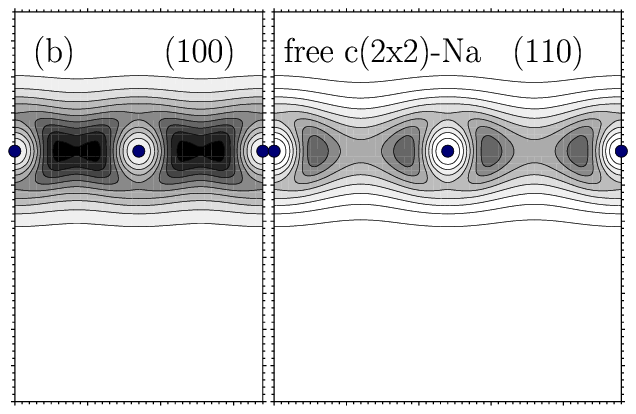,rheight=45mm}
\caption{Free $c(2 \times 2)$-Na layer: (a) band structure (b) charge density
distribution of valence elctrons. The planes of cross-sections are defined in
Fig.~4b.}
\end{figure}

We show in Fig.~9 the total charge
density, $\rho_{\rm tot.}$, the density difference between the adsorption system
and the corresponding Al surface (for which the positions of the 
%
%
\begin{figure}
\vspace{-10mm}
\psfig{figure=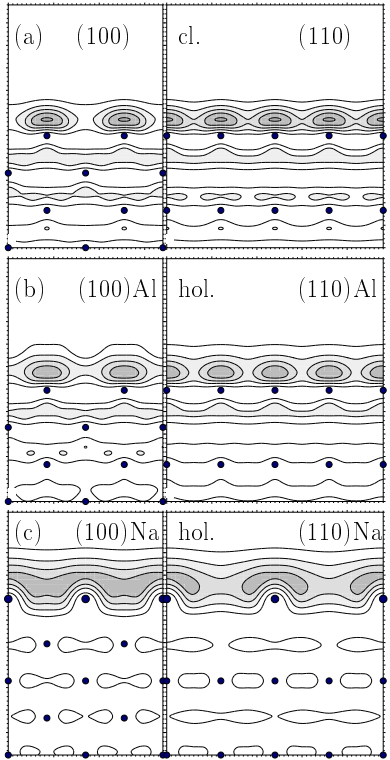,rheight=115mm}
\caption{Charge density distribution of the surface
state/resonance of the main bands at $\overline\Gamma$ for
(a) the clean surface, (b) the Al-derived band of the LT phase (hollow site),
and (c) the Na-derived band of the LT phase. The
positions of cross-sections are indicated in Figs.~4a and 4b.
Large dots denote Na atoms and small dots Al atoms.
The contours begin at 1.0 and the contour spacing is 1.0.
The units are $\times 10^{-3}$ $e$ bohr$^{-3}$. }
\end{figure}
Al atoms are 
kept at those of the adsorption system),
$\Delta \rho_{\rm diff.}=\rho_{\rm tot.}-\rho_{\rm Al}$,
and the {\em redistribution} of charge,
$\Delta \rho_{\rm redis.}= 
\rho_{\rm tot.}-\rho_{\rm Al}-\rho_{\rm c(2 \times 2)-Na}$,
which subtracts out the charge density of the
free Na monolayer (Fig.~5b), showing exactly where charge has been enhanced
and depleted. It can clearly be seen from $\Delta \rho_{\rm diff.}$ and 
$\Delta \rho_{\rm redis.}$ 
that charge enhancement occurs primarily at the maxima of the surface states,
and indeed almost proportionally.
We can also note in $\Delta \rho_{\rm redis.}$ some regions of depletion, 
showing that the
electron charge has been transferred from the upper half of the region between 
the Na atoms, where the density for the free Na layer (see Fig.~5b) is much
larger than that of the Al surface states (see Fig.~6a).
We may thus conclude that the electron transfer is occurring from Na
atoms directly into the {\em pre-existing surface states} of Al.

It is noted in passing that this character of the LT phase shows a close 
analogy to the case of $c(2 \times 2)$-Cs/W\,(001) \cite{wimmer,soukiassian},
both having the $c(2 \times 2)$ four-fold hollow structure. Although the
electronic structure is much more complicated for Cs/W, the essential feature of 
the charge transfer is the same. Thus, the
maxima of the 
%
%
\begin{figure}
\vspace{-10mm}
\psfig{figure=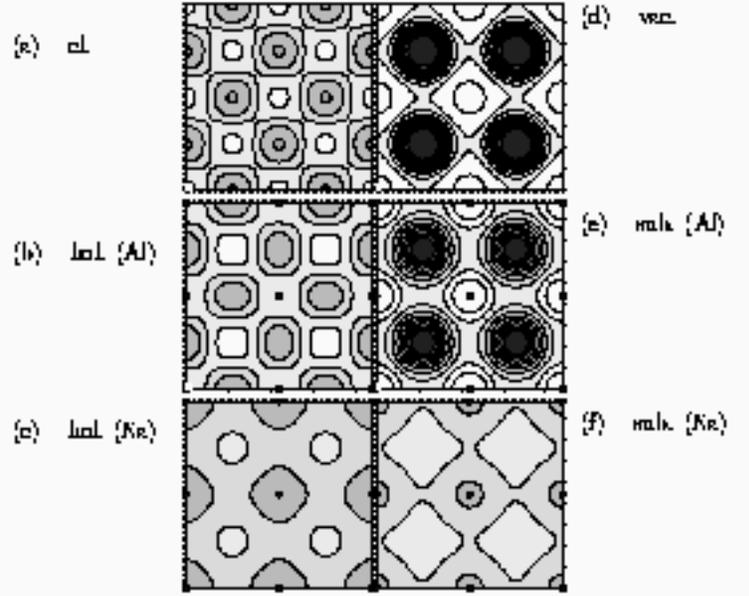,rheight=90mm}
\caption{Charge density distribution of the main surface state/resonances
in a plane
parallel to the surface passing through the electron density
maxima ({\em not} the atom centres):
The clean surface (a), the Al-derived (b)  and Na-derived (c)
bands of the LT (hollow) structure, respectively, and the vacancy structure (d),
the  Al-derived (e)  and Na-derived (f) bands of the RT (substitutional) structu
re, respectively.
The dots represent the positions of the Na atoms.
The contours begin at 1.0 and the contour spacing is 1.0.
The units are $\times 10^{-3}$ $e$ bohr$^{-3}$. }
\end{figure}
4d-derived surface states of the clean W\,(001) surface lie on top
of the surface W atoms (see Fig.~7 of Ref.\cite{wimmer}). The maxima maintain their form upon
Cs adsorption. Charge transfer takes place from Cs to
these surface states. Hence, this case and ours
may be regarded as representatives of the alkali-metal on
metal systems for which the surface states of the substrate play
an essential role.

The upper Na-derived band is crossing the Fermi level and partly occupied, 
remaining ``metallic''. In Figs.~6c and 7c its
free-electron-like character can be seen. We see in Figs.~6b and 6c that the
charge density of the Al-derived band lies well below the
smeared out density of the Na-derived band, and also below the position of 
the Na atoms.
 Thus, the traditional picture of a thin metallic film covering an originally
metallic substrate remains qualitatively valid. We see below that this is 
{\em not} the case for the RT phase.

It is to be noted that the apparently weak influence of the Na adsorption on 
the Al surface states is limited to the high coverage of ${\Theta}$=0.5 for the
$c(2 \times 2)$ structure, for which the density maxima for the Na valence
electrons (Fig.~5b) lie {\em between} the Na atoms and situated just at the same 
site as the surface state maxima. In fact, it has been found experimentally 
\cite{fasel} that at low coverages ${\Theta}$=0 to 0.15 the surface states 
are deteriorated by Na adsorption. Also it has been shown by DFT  
calculation \cite{stampfl-unpub} for a fictive ordered-structure model, 
$p(2 \times 2)$ with coverage ${\Theta}$=1/4, that the surface-state
maxima are moved from the on-top sites to the bonding-line positions between
the Na and Al atoms. 
%
%
\begin{figure}
\psfig{figure=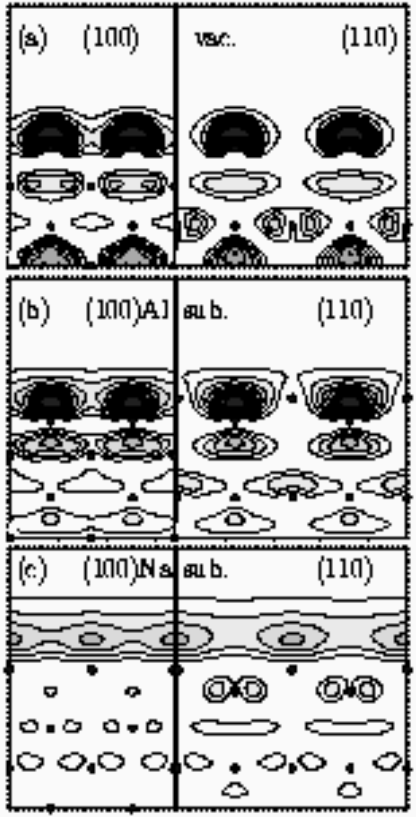,height=120mm}
\caption{Same as Fig.~6 for the RT phase (substitutional site).
The positions of cross-sections are specified in Figs.~4c and 4d.}
\end{figure}

\subsection{RT $c(2 \times 2)$ substitutional structure}

\subsubsection{Al-derived band}

Figure 2c shows the experimental (left panel) and calculated
(right panel) surface state/resonance bands for the RT $c(2 \times 2)$ 
substitutional structure. 
It can be seen that the experimental surface state/resonance band 
clearly exhibits a $c(2 \times 2)$ periodicity,
in contrast to the LT (hollow) structure. 
As we will see below, this is due to the significantly reconstructed Al\,(001) 
surface.
The calculations (right panel) show good agreement with experiment and indicate 
that the band is mainly Al-derived (circles) but has a small participation of Na
(squares). The energy position of the main theoretical and experimental bands 
at $\overline{\Gamma}$ are 2.14 and 2.31~eV, respectively.

The experimental results display some asymmetry in the intensity near
{\bf k}$_{\parallel}$=0.5 in $\overline{\Gamma}$-$\overline{\rm M}$, i.e. a 
weakening on the left side and a strengthening and upward shift on the right.
A few alternative explanations may be possible for this anomaly.
Among others, it may be related to the closing of the bulk band gap in this 
region, causing the state to go from being a pure surface state to a surface 
resonance (see Fig.~10c of Ref.~\cite{fasel}). Also, additional coupling to 
bulk states may become possible by surface Umklapp processes. The peak profile
may become broad and complicated, as indicated by the split structure  of the band
near {\bf k}$_{\parallel}$=0.5 in $\overline{\Gamma}$-$\overline{\rm M}$. 
Different profiles for the left- and right-half may result from 
this broadened peak in the formation of the matrix element given by Eq.~4, 
the final state being different. 

An important finding in the present analysis is that the main band in the RT 
structure can be regarded as being derived from the 
surface state/resonance band of a fictive, reconstructed clean
Al surface, that is, the ``vacancy'' structure, see Fig.~4c. In
this structure the Na atoms are replaced by "vacancies" of the Al atoms.
The
surface state/resonance band structure of the vacancy structure
is displayed in Fig.~3c. 
By comparison with Fig.~3b it is clearly seen that the
main Al-derived band of the RT phase originates from that of the 
vacancy structure and is only somewhat shifted down in energy (by 0.81~eV)
due to Na adsorption.

Figures~8a and 8b show the single-state charge
distribution at $\overline{\Gamma}$ for the
main bands of the  ``vacancy structure"
and the RT phase, respectively.
Figures~7d and 7e show their cross-sections parallel to the surface
passing through the electron density maxima.
Similarly to the clean Al(001) surface,
the maxima of charge density for the vacancy structure
lie on top of the uppermost surface Al
atoms, having this time the $c(2 \times 2)$ periodicity. For the Al-derived band
of the RT phase,  the maxima also lie  approximately at the same
position as that of the vacancy structure with relatively small changes
in their form.  It can be noted that these states are rather strongly localized.
This explains the well developed $c(2 \times 2)$ character of the band found in 
Fig.~2c, and its relatively small dispersion (larger value of m*, see Table II).

Similarly to the LT phase, we find in Fig.~2c a few additional theoretical bulk
bands which are not present for the
clean surface. These are apparently introduced by the $c(2 \times 2)$
periodicity as surface Umklapp processes.
There is also indication of a third, relatively weak band lying at
E$_{\rm F}$ and around $\overline{\Gamma}$
and $\overline{\rm M}$ in both the experimental
and calculated band structures.
From our analysis (not shown), this band
can be regarded as being derived from a second, relatively weak,
surface state/resonance band of the vacancy structure.

\subsubsection{Na-derived band}

In Fig.~3b we see that the Na-derived band (filled squares)
is found only in the calculation, lying completely in the
positive energy range, about 2.7 eV higher in energy than for the
free $c(2 \times 2)$-Na layer. Notably, this shift is much larger than the 
downward shift of the Al-derived band (~0.8 eV). This will be discussed below.

The single-state charge distribution of the unoccupied Na-derived 
state at $\overline{\Gamma}$ is shown in Fig.~8c, indicating again an 
antibonding character. 
Figures~8c and 7f show its smeared out, free-electron-like structure with the
maxima residing this time above the Na atoms.
%
%
\begin{figure}
\vspace{-10mm}
\psfig{figure=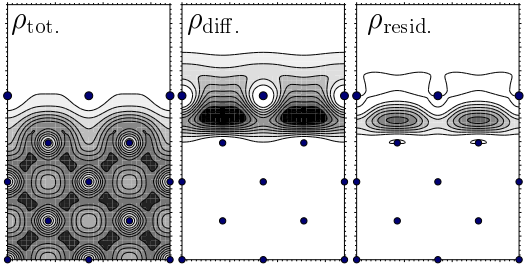,rheight=50mm}
\caption{Total charge density distribution, $\rho_{\rm tot.}$, (left panel),
density difference, $\Delta \rho_{\rm diff.}$, (middle panel),
and density redistribution, $\Delta \rho_{\rm redis.}$, (right panel),
of the LT phase (hollow site).
The cross-section is in the (001)Na plane (see Fig.~4b).
The units are $10^{-3}$ $e$ bohr$^{-3}$.
In the left panel the first contour begins
at 4.0 with a spacing of 4.0, for the  middle panel the first contour
begins at 0.6 with a spacing of 0.6, and in the right panel the contours are
the same as the middle panel with the addition of a negative contour line
(unshaded) at $-$0.6. }
\end{figure}

\subsubsection{Formation mechanism of the two bands}

We conclude that the two main 
bands are resulting, just as in the case of the LT phase,
from the coupling between the states of a free $c(2 \times 2)$-Na
monolayer and the surface state/resonance of the Al surface,
i.e., the vacancy structure in this case.

The electron transfer can also be regarded as occurring from Na
atoms into the surface state/resonance of the vacancy structure. This can be
seen clearly in comparing Fig.~10 with Fig.~8a.
In the charge redistribution $\Delta \rho_{\rm redis.}$ in Fig.~10 we note 
some regions of depletion: the
electron charge has been transferred in this case mainly from the region on top 
of the Na atoms,
where the density of the free Na layer dominates the density of the
surface state/resonance of the vacancy structure. 
The charge depletion found in Fig.~10 also near the centres of the Al
atoms may be interpreted as a result of an upward shift of the maximum of the
surface states induced by Na adsorption.
%
%
\begin{figure}
\vspace{-10mm}
\psfig{figure=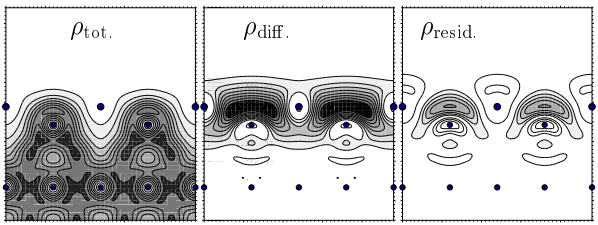,rheight=40mm}
\caption{Same as Fig.~9 for the RT phase (substitutional site).
For the  middle panel there is one contour line at $-1.2$,
and in the right panel there are three negative contour lines
(unshaded) at $-$1.8, $-$1.2, and $-$0.6.}
\end{figure}

These findings, combined with the unbalancedly large upward shift of the 
Na-derived band and the existence of the third peak at E$_{\rm F}$, both mentioned 
above, indicate that the coupling between the free Na monolayer states
and the Al surface states/resonances takes place in a little more complicated
way than in the LT case. It can no longer be interpreted in a simple 
two-term-coupling perturbation scheme. Apparently the coupling is too strong, and 
involves other states, namely, the $p_{z}$ states of Na lying originally at +1 eV
(Fig.~5a) and the second surface states/resonances 
of Al lying originally at ~0 eV (Fig.~3c).

We have seen above that for the LT phase the traditional picture of a thin 
metallic film covering a metallic substrate remains qualitatively valid.
For the RT phase, on the other hand, we see in Fig.~8b that the completely filled
Al-derived band has the maxima lying at the {\em same} height as the Na atoms.
The maxima of the Na-derived band lie indeed higher, but they are empty.
Thus, as far as the electronic states in the energy range near E$_{\rm F}$ are 
concerned, 
the Al-derived band constitutes practically the "surface" of this structure.
The Na monolayer (or the Na/Al composite monolayer) cannot 
be regarded as a metallic film on a metallic substrate; rather,
except for the still existent background bulk-continuum,
the  electronic structure of the monolayer may be viewed
as being analogous to that of an {\em ionic} crystal,
like NaCl, where the completely filled Al-derived band replaces the valence band
of Cl, and forms with the empty Na-derived band a "band-gap" of about 3.5 eV
across the Fermi energy.

We note in passing that we find a similarity of the surface electronic structure
of the RT phase to that of the system
$(2 \times 2)$-2Na/Al\,(111) which forms a composite double layer
surface alloy with a similar, but more complex intermixing of Na
and Al in the surface layers \cite{stam1,stam2}.
For this system we find also a filled Al-derived band originating from the 
corresponding ``vacancy structure'', and an empty Na-derived band. (The latter may be 
assigned to that found by Heskett {\em et al.}~ \cite{heskett} using inverse 
photoemission spectroscopy.)
The similarity indicates that also this layer cannot be regarded simply as a thin
metallic layer.

\section{Discussion}

For all three systems studied (Al\,(001) and the two
 $c(2 \times 2)$-Na/Al(001) phases),
we find good over-all agreement of the surface state/resonance band
structure between experiment and theory.
In each case there is a prominent Al-derived surface-state/resonance band,
showing similar free-electron-like parabolic dispersion at the band bottom
near $\overline{\Gamma}$, as indicated by broken lines in Fig.~2.
We find that the observed and calculated values
of the Fermi wave vector, {\bf k}$_{\rm F}$, and the effective mass, $m^{*}$
of the main band (Tab.~II) agree only roughly, probably due to the approximations 
used in calculation, as already mentioned.  In any case these
quantities are to be regarded as global parameters specifying only the geometry
of the bands, and are not intended to
indicate that the bands are free-electron-like. In fact, the
wave functions of the Al-derived bands are rather strongly localized, as we see 
from the charge density
distributions. The dispersion of the bands may be regarded as the result of the
overlap between the localized wave functions.

The picture of charge transfer taking place from Na to the surface
state/resonances of Al is also supported by the almost equal values of 
the work function change (decrease) 
$\Delta \Phi \approx $-$1.6$~eV, obtained both
experimentally~\cite{porteus,paul} and theoretically in the
present work for the LT and RT phases.
Obviously the value of $\Delta \Phi$ results from $\rho_{\rm redis.}$,
shown in Figs.~9 and 10, as the $z$-component of 
the dipole moment. As already mentioned, the {\em minimum}
and maxima of $\rho_{\rm redis.}$ correspond to the maxima
of the Na- and Al-derived states, respectively.
This can be seen by comparison of Figs.~9 and 10 with Figs.~6b,c 
and Figs.~8b,c.
In these figures, the vertical distances between the maxima
of the Na- and Al-derived states are found to have the ratio of about 1.0 
to 0.6 between the LT and RT phases. On the other hand, 
a comparison between Figs.~5a and 2b,c shows that  the decrease of
the occupancy of the
Na-derived states amounts to ca. 0.6 for the LT phase, in contrast 
to 1.0 for the RT phase (being empty).
This lead to the same amount of $\Delta \Phi$  between the two phases, as
a product of charge and distance. It is essential in this
consideration to note the fact that, while the forms of the 
Al-derived states remain always almost  unchanged, the forms of the
Na-derived states change drastically, from that of Figs.~5b to
those of Figs.~6c and 8c.

Our results verify the formation of
the {\em two} surface state/resonance bands. Various experimental methods
of studying surface electronic properties other than ultraviolet
photoelecton spectroscopy, as used here, may be useful for
finding out characteristic features induced by these two
bands. Particularly, EELS (electron energy loss spectroscopy)
\cite{aruga,unocc}
and  IPS (inverse photoemission spectroscopy) 
\cite{heskett,unocc}
may be useful for studying the effects of the bands lying in
the energy range above the Fermi level. 

For the RT phase, the fact that the surface band structure is constituted  of
a filled band and an empty band with a gap of  
approximately 3.5~eV, 
may play an important role in various properties
of the surface. We may think of, in addition to EELS, 
an anomalous feature in optical reflection spectra \cite{mcintyre} near 3.5~eV
and a corresponding anomaly in the dispersion curve of surface plasmons
\cite{raether}. 

A significant effect may be expected in various
surface-sensitive methods, such as 
ion neutralization \cite{hagstrum}, metastable
deexcitation \cite{hagstrum,kueppers} etc.,
which would reflect the dominance of the surface states/resonances
in the outermost surface region, and hence expose the difference in the 
character of these states between the LT and RT phases.
We may also think of the relevance of the occupancy of surface 
states/resonances on surface diffusion \cite{bertel} and catalytic activity 
\cite{memel}.

\section{Conclusion}

We have analysed the electronic structure
of the metastable hollow and stable substitutional 
structures  of $c(2 \times 2)$-Na on Al\,(001).
The calculated surface state/resonance bands agree well with those
measured by angle-resolved photoemission experiments.
It is found that in both
phases, two pronounced bands appear as the  result of
a characteristic coupling between the
valence-state bands of a free $c(2 \times 2)$-Na layer and the
surface-state/resonance bands of the corresponding (i.e. clean and reconstructed)
Al surfaces. 
While the experimental band structure of  the substitutional
structure shows a clear $c(2 \times 2)$ character
due to the significant reconstruction of the
surface Al layer, the hollow structure does not:
The main band, in fact, exhibits a quasi-$(1 \times 1)$ periodicity
like that of the clean surface.
For the stable substitutional structure, the unoccupied surface
state/resonance band lies completely above the Fermi level,
leading to the formation
of a surface-state band structure that resembles that of an
ionic crystal. We await experimental verification of the 
predicted unoccupied surface state/resonance band, and of the difference
in the properties of the LT and RT phases in relation to the character of the
surface states/resonances.

\begin{table}
\begin{tabular}{ccccc}
\multicolumn{3}{c}{} & \multicolumn{2}{c}{ Min. Fraction} \\
Structure  & Symbol & Projected Atom & $\overline{\Gamma}$-$\overline{\rm X}$ & 
$\overline{\Gamma}$-$\overline{\rm M}$\\
\hline
clean   & big circles & Al  & 0.8 & 0.8 \\
        & small circles & Al  & 0.4 & 0.4 \\
\hline
LT      & empty squares  & Na  & 0.5 & 0.5 \\
        & big circles & Al  & 0.7 & 0.8 \\
        & small circles & Al  & 0.35 & 0.4 \\
\hline
RT      & big squares  & Na  & 0.4 & 0.5 \\
        & small squares  & Na  & 0.2 & 0.2 \\
        & small circles & Al  & 0.5 & 0.55 \\
\hline
vacancy & big circles & Al  & 0.5 & 0.5 \\
        & small circles & Al  & 0.3 & 0.3 \\
\end{tabular}
\caption{Legends of the symbols used in calculated 2D band structures
in Figs.~2 and 3. The "minimum fraction" is defined in the text (see III
Calculation method).}
\end{table}

\begin{table}
\begin{tabular}{ccccc}
Structure & $E_{0}$[eV] & $k_{\rm F}[\frac{2\pi}{a_{0}}]$ & $m^{*}[m_{e}]$ \\  
\hline
clean     & 2.68 (2.76) & 0.55 (0.60) & 1.05 (1.18) \\
LT        & 3.61 (3.12) & 0.64 (0.66) & 1.03 (1.29) \\
RT        & 2.14 (2.31) & 0.51 (0.62) & 1.12 (1.55) \\
vacancy   & 1.33        & 0.41        & 1.15        \\
\end{tabular}
\caption{Parameters specifying the main surface state/resonance
bands in Figs.~2 and 3. The experimental values [10] are shown in brackets.}
\end{table}


\begin{references}
\bibitem{osterwalder2} Details about the determination and
calibration of Na coverages can be found in Ref.~[10].
\end{references}
\end{document}